\documentclass[letterpaper,11pt]{article}
\usepackage{url}
\usepackage{hyperref}
\usepackage{graphicx}
\usepackage{amsmath}
\usepackage[affil-it]{authblk}
\newcommand{\nue}{\mbox{$\nu_e$}}
\newcommand{\nuebar}{\mbox{$\overline\nu_e$}}
\newcommand{\numu}{\mbox{$\nu_{\mu}$}}
\newcommand{\numubar}{\mbox{$\overline\nu_{\mu}$}}
\newcommand{\nutau}{\mbox{$\nu_{\tau}$}}
\newcommand{\nutaubar}{\mbox{$\overline\nu_{\tau}$}}

\newcommand{\thetatwothree}{\mbox{$\theta_{23}$}}

\newcommand{\dcp}{\mbox{$\delta_{CP}$}}

\newcommand{\sigmahat}{\mbox{$\hat{\sigma}$}}

\makeatletter
\def\and{%                  % \begin{tabular}[t]{c}
  \end{tabular}%
  \hskip 0.01em %
  \begin{tabular}[t]{c}}%   % \end{tabular}
\makeatother

\title{Experiment Simulation Configurations Used in DUNE CDR}
\date{}

\author[14]{T.~Alion}
\author[15]{J.~J.~Back}
\author[8]{A.~Bashyal}
\author[13]{M.~Bass}
\author[2]{M.~Bishai}
\author[3]{D.~Cherdack}
\author[2]{M.~Diwan}
\author[1]{Z.~Djurcic}
\author[11]{J.~Evans}
\author[7]{E.~Fernandez-Martinez}
\author[4]{L.~Fields}
\author[16]{B.~Fleming}
\author[12]{R.~Gran}
\author[13]{R.~Guenette}
\author[11]{J.~Hewes}
\author[3]{M.~Hogan}
\author[4]{J.~Hylen}
\author[4]{T.~Junk}
\author[9]{S.~Kohn}
\author[4]{P.~LeBrun}
\author[4]{B.~Lundberg}
\author[4]{A.~Marchionni}
\author[10]{C.~Morris}
\author[4]{V.~Papadimitriou}
\author[4]{R.~Rameika}
\author[4]{R.~Rucinski}
\author[11]{S.~S\"oldner-Rembold}
\author[6]{M.~Sorel}
\author[5]{J.~Urheim}
\author[2]{B.~Viren}
\author[10]{L.~Whitehead}
\author[3]{R.~Wilson}
\author[2]{E.~Worcester}
\author[4]{G.~Zeller} 
\affil[1]{Argonne National Lab., Argonne, IL 60439, USA}
\affil[2]{Brookhaven National Lab., Upton, NY 11973-5000, USA}
\affil[3]{Colorado State University, Fort Collins, CO 80523, USA}
\affil[4]{Fermi National Accelerator Lab, Batavia, IL 60510-0500, USA}
\affil[5]{Indiana University, Bloomington, IN 47405-7105, USA}
\affil[6]{Instituto de Fisica Corpuscular, C/Catedratico Jose Beltran, 2
  E-46980 Paterna (Valencia), Spain}
\affil[7]{Madrid Autonoma University, Ciudad Universitaria de Cantoblanco
  28049 Madrid SPAIN, Spain}
\affil[8]{Oregon State University, Dept. of Physics;
  301 Weniger Hall; Corvallis, OR 97331\textendash6507, USA}
\affil[9]{University of California (Berkeley), \#7300;
Berkeley, CA 94720-7300, USA}
\affil[10]{University of Houston, Houston, TX 77204, USA}
\affil[11]{University of Manchester, Oxford Road, Manchester M13 9PL, UK}
\affil[12]{University of Minnesota (Duluth), Duluth, MN 55812, USA}
\affil[13]{University of Oxford, Oxford, OX1 3RH, UK}
\affil[14]{University of South Carolina, Columbia, SC 29208, USA}
\affil[15]{University of Warwick, Coventry CV4 7AL, UK}
\affil[16]{Yale University, New Haven, CT 06520, USA}
\begin{document}

\maketitle

\begin{abstract}
The LBNF/DUNE CDR describes the proposed physics program and experimental design
at the conceptual design phase. Volume 2, entitled \textit{The Physics
Program for DUNE at LBNF}, outlines the scientific objectives and describes the
physics studies that the DUNE collaboration will perform to address these objectives.
The long-baseline physics sensitivity calculations presented in the DUNE CDR rely upon
simulation of the neutrino beam line, simulation of neutrino interactions in the far
detector, and a parameterized analysis of detector performance and systematic
uncertainty. The purpose of this posting is to provide the results of these
simulations to the community to facilitate phenomenological studies of long-baseline
oscillation at LBNF/DUNE. Additionally, this posting includes GDML of the DUNE single-phase
far detector for use in simulations. DUNE welcomes those interested in
performing this work as members of the collaboration, but also recognizes the benefit of making these
configurations readily available to the wider community. 
\end{abstract}

\section{Introduction}
The Conceptual Design Report (CDR) for the Long-Baseline Neutrino Facility (LBNF) and the Deep Underground Neutrino
Experiment (DUNE) describes, in three volumes -- Volume~1: The LBNF and DUNE Projects\cite{cdrv1},
Volume~2: The Physics Program for DUNE at LBNF\cite{cdrv2}, and Volume~4: The DUNE Detectors at LBNF\cite{cdrv4},
the design and proposed physics program for LBNF/DUNE.
The primary scientific objectives of LBNF/DUNE are
to study long-baseline neutrino oscillation to determine the neutrino mass ordering, to determine whether
CP symmetry is violated in the lepton sector, and to precisely measure the parameters governing neutrino
oscillation to test the three-neutrino paradigm. The DUNE physics program also includes precise measurements
of neutrino interactions, observation of atmospheric neutrinos, searches for nucleon decay, and sensitivity
to supernova burst neutrinos.
LBNF consists of the technical and conventional facilities for a high-power neutrino beam, civil construction of
the near detector facility, and excavation and underground infrastructure for the far detector caverns. DUNE
consists of the near detector systems and four liquid argon TPC (LArTPC)
far detector modules, each with a fiducial mass of about 10 kt.

In Volume 2 of the CDR, the proposed physics program for DUNE is presented. The long-baseline physics sensitivity
calculations presented in Volume 2 are based upon detailed predictions for the expected neutrino flux,
kinematics of neutrino interactions in the far detector, parameterized simulations of detector performance, realistic
event selection criteria, and uncertainty in signal normalization as an approximation of the effect of systematic uncertainties.
This posting provides the results of these simulations for use by anyone in the community interested in studying long-baseline
neutrino oscillation. The text in this document is not intended to provide thorough documentation of the details of how these
results are produced; rather we attempt to briefly summarize the analyses that produce these results and provide documentation
of how the results may be used.

In Section~\ref{sect:flux}, we describe the simulated LBNF fluxes for a reference and optimized beam
design, at both the near and far detectors, in both forward-horn current (FHC) and reverse-horn current (RHC) modes, provided in the
ancillary files in directory DUNE\_Flux/.
In Section~\ref{sect:fastmc}, we describe simulation and analysis of the expected event samples in the Far Detector
using the Fast MC. The results of this analysis are provided in the ancillary files in directory
DUNE\_GLoBES\_Configs/, containing a GLoBES\cite{globes1,globes2} configuration, which is described in Section~\ref{sect:globes}.
In Section~\ref{sect:gdml}, we describe the
far detector geometry, several versions of which are provided in the ancillary
files in directory DUNE\_GDML/.

\section{Flux Simulation}
\label{sect:flux}
The neutrino fluxes used in the CDR were produced using G4LBNF, a Geant4\cite{GEANT4:NIM,GEANT4}-based simulation of the LBNF beamline
from primary proton beam to hadron absorber. Specifically, G4LBNF version v3r3p6 was used, which was built against Geant4
version 4.9.6p02.  All simulations used the QGSP\_BERT physics list.

\begin{figure}[!htpb]
  \centering
  \includegraphics[width=0.8\textwidth]{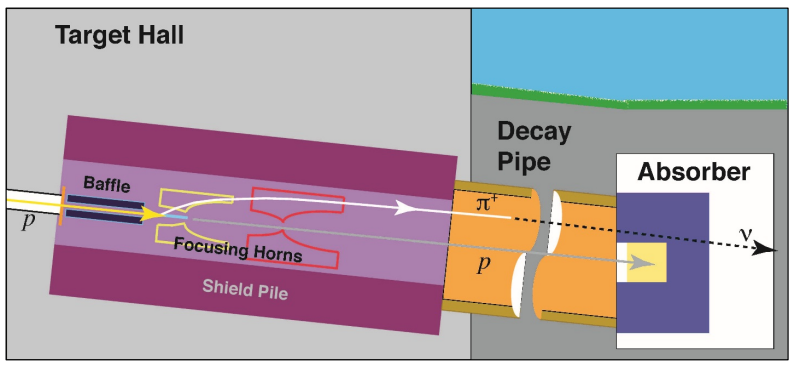}
  \caption{Cartoon of the neutrino beamline showing the major components of the neutrino beam: the beam window, horn-protection baffle,
  target, toroidal focusing horns, decay pipe, and absorber. Figure from \cite{cdr_annex3a}.}
  \label{fig:beam_cartoon}
\end{figure}

G4LBNF is highly configurable to facilitate studies of a variety of beam options.  For the fluxes provided here, it was configured
to simulate the reference beam described in detail in Annex 3A of the CDR\cite{cdr_annex3a},
and the ``optimized'' design summarized in section 3.7.2
of Volume 2\cite{cdrv2}.
The reference beam design is based on the design of the target and focusing system for NuMI\cite{Anderson:1998zza}.
The baffle is a 1.5-m
graphite cylinder, which protects downstream equipment in the case of a mis-steered beam. The graphite target is 95~cm
long, corresponding to two interaction lengths. The target is surrounded by the first horn and followed by the second horn, each of which
has a parabolic geometry and is operated at a current of 230 kA. The 194-m decay pipe is filled with helium and is followed by an
absorber. A cartoon of the neutrino beamline is shown in Fig.~\ref{fig:beam_cartoon}. In the optimized design, a genetic algorithm
is employed to determine values for 20 beamline parameters describing the primary proton momentum, target dimensions, and horn
shapes, positions, and current that maximize sensitivity to CP violation in DUNE. Further optimization of the beam design following
this approach is ongoing in the collaboration.

Both the optimized and reference geometries include a detailed description of the target, baffle, decay pipe,
hadron absorber, and shielding.  The reference geometry also includes a detailed description of both focusing horns, including welds
and spider support.  The optimized geometry uses a simplified model of the focusing horns.
The basic output of G4LBNF is a list of all particle decays to neutrinos that occur anywhere along the beamline.
Weights (historically referred to as “importance weights”) are used to reduce the size of the output files by throwing out
a fraction of the relatively common low-energy neutrinos while preserving less numerous high-energy neutrinos.
To produce neutrino flux distributions at a particular
location, all of the neutrinos in the G4LBNF output file are forced to point toward the specified location and weighted according
to the relative probability that the decay in question would produce a neutrino in that direction\cite{pavlovic_thesis}.

For each beam option, fluxes are provided at the center of the near detector (ND), located 459~m downstream of the start of Horn 1, and
at the far detector (FD), located 1297~km downstream of the start of Horn 1.  Fluxes are available for both neutrino mode (FHC) and
antineutrino mode (RHC).  Each flux is available in two formats: a root file containing flux histograms and a
GLoBES flux input file.  The root files also contain neutral-current and charged-current spectra,
which are obtained by multiplying the flux by GENIE 2.8.4
inclusive cross sections. The flux histograms in the root files have units of neutrinos/$\mathrm{m^2}$/POT.
Note that these histograms have variable bin widths, so discontinuities in the number of events per bin
are expected.
The GLoBES files have units of neutrinos/GeV/$\mathrm{m^2}$/POT. These text files are in the standard GLoBES
format, in which the seven columns correspond to: $E_{\nu}, \Phi_{\nue}, \Phi_{\numu}, \Phi_{\nutau}, \Phi_{\nuebar}, \Phi_{\numubar},
\mathrm{and} \, \Phi_{\nutaubar}$.

\section{Fast MC Simulation}
\label{sect:fastmc}
As described in the DUNE CDR,
the LArTPC performance parameters that go into the sensitivity calculations
are generated using the DUNE Fast Monte Carlo (MC) simulation, which
is described in detail in~\cite{Adams:2013qkq}.  The Fast MC combines
the simulated flux, the GENIE neutrino interaction
generator~\cite{Andreopoulos:2009rq}, and a parameterized detector
response that is used to simulate the reconstructed energy and
momentum of each final-state particle. The detector response
parameters used to determine the reconstructed quantities are
summarized in Table~\ref{tab:fastmc_detectorinputs}. The assumptions
on detector response used in the Fast MC are preliminary, and are
expected to improve as the full detector simulation advances and
more information on the performance of LArTPC detectors becomes
available.
The simulated energy
deposition of the particles in each interaction is then used to
calculate reconstructed kinematic quantities (e.g., the neutrino
energy). Event sample classifications ($\nu_e$ CC-like, $\nu_{\mu}$
CC-like, or NC-like), including mis-identification rates, are determined by the
identification of lepton candidates. Lepton candidates are selected
based on a variety of criteria including particle kinematics,
detector thresholds, and probabilistic estimates of particle interaction
final states. To
reduce the neutral-current (NC) and $\nu_{\tau}$ charged-current (CC)
backgrounds in the $\nue$ and $\numu$
CC-like samples, additional discriminants are formed using reconstructed
transverse momentum along with reconstructed neutrino and hadronic
energy as inputs to a k-Nearest-Neighbor (kNN) machine-learning
algorithm. Plots showing the true-to-reconstructed smearing matrices, the analysis
selection efficiencies, and the expected far detector event spectra 
generated by the Fast MC are available in \cite{cdrv2}.

\begin{table}[!htb]
  \centering
  \caption{Summary of the single-particle
    far detector response used in the Fast MC. For some particles, the response depends upon behavior or
    momentum, as noted in the table. If a muon or a pion that is mis-identified as a muon is
    contained within the detector, the momentum is smeared based on track length.
    Exiting particles are smeared based on the contained energy.
    For neutrons with momentum less than 1~GeV/c,
    there is a 10\% probability that the particle will escape detection, in which case the reconstructed energy is
    set to zero. For neutrons that are detected, the reconstructed energy is taken to be 60\%
    of the deposited energy after smearing.}
  \label{tab:fastmc_detectorinputs}
  \begin{tabular}{|l|c|c|c|} \hline
    Particle type & Detection & Energy/Momentum & Angular \\ 
    & Threshold (KE) & Resolution & Resolution \\ \hline\hline
    $\mu^{\pm}$ & 30 MeV & Contained track: track length  & 1$^\circ$\\
    &        & Exiting track: 30\%            & \\ \hline
    $\pi^{\pm}$ & 100 MeV & $\mu$-like contained track:  track length & 1$^\circ$\\
    &         & $\pi$-like contained track: 5\% & \\
    &           & Showering or exiting: 30\% & \\ \hline
    e$^{\pm}$/$\gamma$ & 30 MeV & 2\% $\oplus$ 15\%/$\sqrt{E}$[GeV] & 1$^\circ$ \\ \hline
    p & 50 MeV & p$<$400 MeV/c: 10\% & 5$^\circ$ \\
    &        & p$>$400 MeV/c: 5\% $\oplus$ 30\%/$\sqrt{E}$[GeV] & \\ \hline
    n & 50 MeV & 40\%/$\sqrt{E}$[GeV] & 5$^\circ$ \\ \hline
    other & 50 MeV & 5\% $\oplus$ 30\%/$\sqrt{E}$[GeV] & 5$^\circ$ \\ \hline
  \end{tabular}
\end{table}

\section{GLoBES Configuration}
\label{sect:globes}
The GLoBES configuration summarizing the result of the Fast MC analysis and facilitating user-generated sensitivities is provided in
the ancillary files in directory DUNE\_GLoBES\_Configs/; the flux included in this configuration is for the Optimized Beam described in
Section~\ref{sect:flux},
but it is valid to substitute the Reference Beam leaving the rest of the configuration unchanged. The flux normalization factor
is included in GLoBES AEDL file to ensure that all variables have the proper units; its value is @norm=1.017718e17.
Cross-section files describing charged-current and neutral-current interactions with argon, generated using GENIE 2.8.4, are included in the configuration.
These cross-section text files are in the standard GLoBES format, in which the seven columns correspond to:
$log_{10}E_{\nu}, \sigmahat_{\nue}, \sigmahat_{\numu}, \sigmahat_{\nutau}, \sigmahat_{\nuebar}, \sigmahat_{\numubar}, \mathrm{and} \,
\sigmahat_{\nutaubar}$,
where $\sigmahat(E) = \sigma(E)/E[10^{-38}\frac{\mathrm{cm}^2}{\mathrm{GeV}}]$.
The true-to-reconstructed smearing
matrices and selection efficiency as a function of energy produced by the Fast MC for various signal and background modes used by 
GLoBES are included. The naming convention for the channels defined in these files is summarized in Table~\ref{tab:fastmc_naming_convention}.

\begin{table}[!htb]
  \centering
  \caption{Description of naming convention for channels included in the GLoBES configuration provided in the ancillary files.
    ``FHC'' and ``RHC'' appear at the beginning of each channel name and
    refer to ``Forward Horn Current'' and ``Reverse Horn Current'' as described in Section~\ref{sect:flux}. Efficiencies are provided
  for both the appearance mode and disappearance mode analyses.}
  \label{tab:fastmc_naming_convention}
  \begin{tabular}{|lcl|} \hline
    Name Includes & Process & Description \\ \hline\hline
    \multicolumn{3}{|l|}{Appearance Mode:} \\
      app\_osc\_nue & $\numu\rightarrow\nue$ (CC) & Electron Neutrino Appearance Signal \\
      app\_osc\_nuebar & $\numubar\rightarrow\nuebar$ (CC) & Electron Antineutrino Appearance Signal\\
      app\_bkg\_nue & $\nue\rightarrow\nue$ (CC) & Intrinsic Beam Electron Neutrino Background\\
      app\_bkg\_nuebar & $\nuebar\rightarrow\nuebar$ (CC) & Intrinsic Beam Electron Antineutrino Background\\
      app\_bkg\_numu & $\numu\rightarrow\numu$ (CC) & Muon Neutrino Charged-Current Background \\
      app\_bkg\_numubar & $\numubar\rightarrow\numubar$ (CC) & Muon Antineutrino Charged-Current Background \\
      app\_bkg\_nutau & $\numu\rightarrow\nutau$ (CC) & Tau Neutrino Appearance Background \\
      app\_bkg\_nutaubar & $\numubar\rightarrow\nutaubar$ (CC) & Tau Antineutrino Appearance Background \\ 
      app\_bkg\_nuNC & $\numu/\nue\rightarrow$~X (NC) & Neutrino Neutral Current Background \\
      app\_bkg\_nubarNC & $\numubar/\nuebar\rightarrow$~X (NC) & Antineutrino Neutral Current Background \\ \hline

      \multicolumn{3}{|l|}{Disappearance Mode:} \\      
      dis\_bkg\_numu & $\numu\rightarrow\numu$ (CC) & Muon Neutrino Charged-Current Signal\\
      dis\_bkg\_numubar & $\numubar\rightarrow\numubar$ (CC) & Muon Antineutrino Charged-Current Signal\\
      dis\_bkg\_nutau & $\numu\rightarrow\nutau$ (CC) & Tau Neutrino Appearance Background \\
      dis\_bkg\_nutaubar & $\numubar\rightarrow\nutaubar$ (CC) & Tau Antineutrino Appearance Background \\
      dis\_bkg\_nuNC & $\numu/\nue\rightarrow$~X (NC) & Neutrino Neutral Current Background \\
      dis\_bkg\_nubarNC & $\numubar/\nuebar\rightarrow$~X (NC) & Antineutrino Neutral Current Background \\ \hline \hline
  \end{tabular}
\end{table}

The GLoBES configuration provided in the ancillary files corresponds to 300 kt-MW-years of exposure: 
3.5 years each of running in neutrino (FHC) and antineutrino (RHC)
mode with a 40-kt fiducial mass far detector, in an 80-GeV, 1.07 MW beam. The $\nue$ and $\nuebar$ signal modes have independent normalization 
uncertainties of 2\% each, while the $\numu$ and $\numubar$ signal modes have independent normalization uncertainties of 5\%.
The background normalization uncertainties range from 5\% to 20\% and
include correlations among various sources of background; the correlations among the background normalization
parameters can be seen by looking at the @sys\_on\_multiex\_errors\_bg parameters in the AEDL file.
The choices for signal and background normalization uncertainties
may be customized by changing the parameter values in the file definitions.inc.
The treatment of correlation among uncertainties in this configuration
requires use of GLoBES version 3.2.16, available from the GLoBES website\cite{globesweb}.

\begin{figure}[!htpb]
  \centering
  \includegraphics[width=0.45\textwidth]{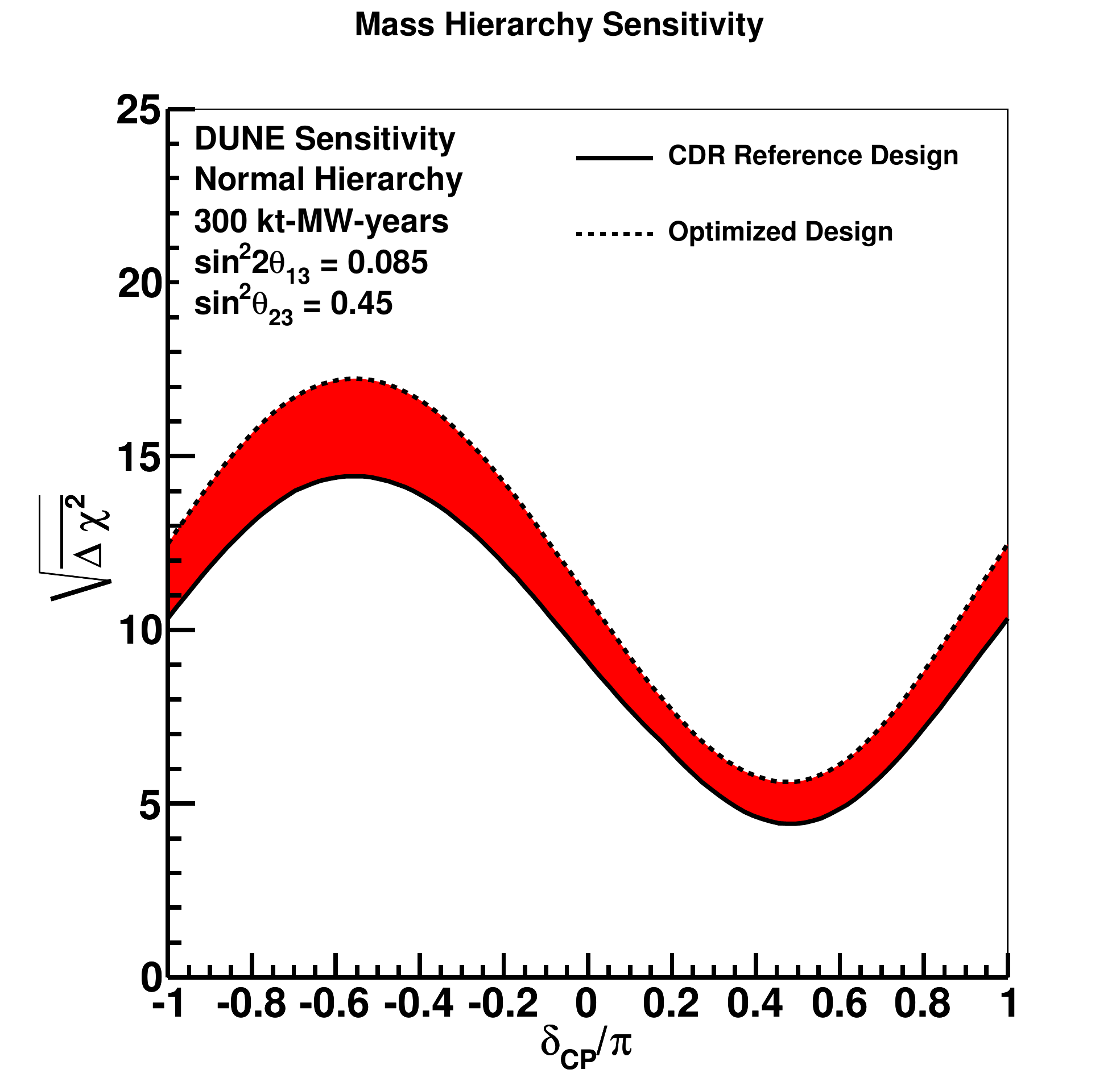}
    \includegraphics[width=0.45\textwidth]{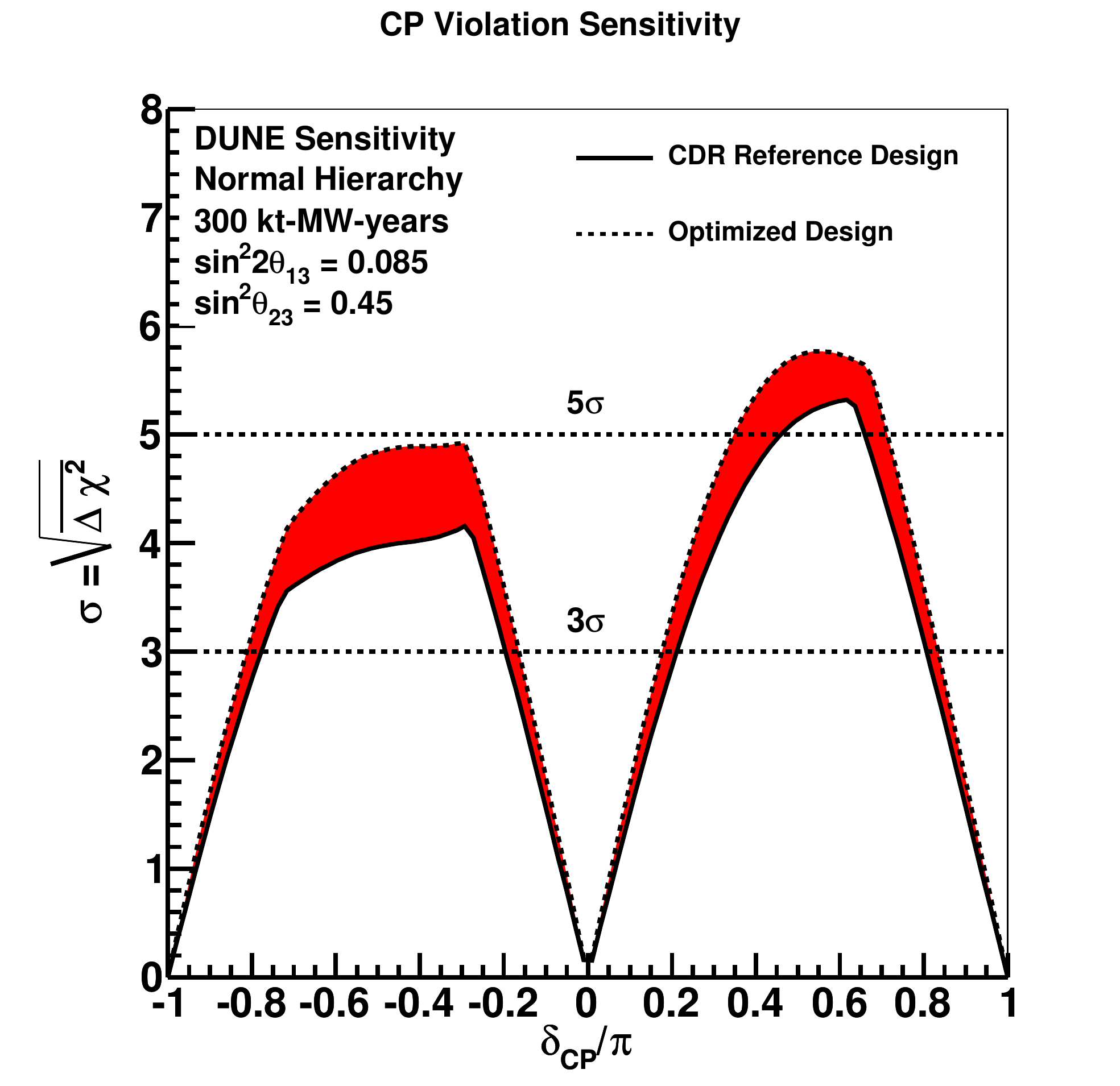}
    \caption{The significance with which the mass hierarchy can be determined (left) or CP violation can be discovered (right)
      as a function of the value of $\dcp$ for an exposure of 300 kt-MW-years, assuming equal exposure in neutrino and antineutrino
      mode and true normal hierarchy. The
      shaded region represents the range in sensitivity due to potential variations in the beam design. Figure from \cite{cdrv2}.}
  \label{fig:osc_sens}
\end{figure}

The sensitivity calculations presented in the CDR use oscillation parameters and uncertainties based on the NuFit
2014\cite{Gonzalez-Garcia:2014bfa} fit
to global neutrino data.
These central values and relative uncertainties are provided in Table~\ref{tab:oscpar_nufit}.
In all cases, oscillation parameters are allowed to vary in the sensitivity calculations, constrained by Gaussian prior functions.
The matter density is constant and equal to the average matter density for this baseline
from the PREM\cite{prem1,prem2} onion shell model of the earth;
the uncertainty on the density is taken to be 2\%.
The GLoBES minimization is performed over both possible values for the $\thetatwothree$ octant and, in the case of CP violation
sensitivity, both possible values for the neutrino mass hierarchy.
Figure \ref{fig:osc_sens} shows the DUNE sensitivity to determination
of the neutrino mass hierarchy and discovery of CP violation, based on the configurations provided here, assuming an exposure of
300 kt-MW-years.

\begin{table}[!htb]
\centering
\caption{Central value and relative uncertainty of neutrino oscillation parameters from a global fit~\cite{Gonzalez-Garcia:2014bfa} to neutrino oscillation data. Because the probability distributions are somewhat non-Gaussian (particularly for $\theta_{23}$), the relative uncertainty is computed using 1/6 of the $\pm3\sigma$ allowed range from the fit, rather than the 1$\sigma$ range.   For $\theta_{23}$ and $\Delta m^2_{31}$, the best-fit values and uncertainties depend on whether normal mass hierarchy (NH) or inverted mass hierarchy (IH) is assumed.}
\label{tab:oscpar_nufit}
\begin{tabular}{|lcc|} \hline \hline
Parameter &    Central Value & Relative Uncertainty \\ \hline
$\theta_{12}$ & 0.5843 & 2.3\% \\ 
$\theta_{23}$ (NH) & 0.738  & 5.9\% \\ 
$\theta_{23}$ (IH) & 0.864  & 4.9\% \\ 
$\theta_{13}$ & 0.148  & 2.5\% \\ 
$\Delta m^2_{21}$ & 7.5$\times10^{-5}$~eV$^2$ & 2.4\% \\ 
$\Delta m^2_{31}$ (NH) & 2.457$\times10^{-3}$~eV$^2$ &  2.0\% \\ 
$\Delta m^2_{31}$ (IH) & -2.449$\times10^{-3}$~eV$^2$ &  1.9\% \\ \hline \hline
\end{tabular}
\end{table}

\section{Far Detector GDML}
\label{sect:gdml}

\begin{figure}[!htpb]
  \centering
  \includegraphics[width=0.8\textwidth]{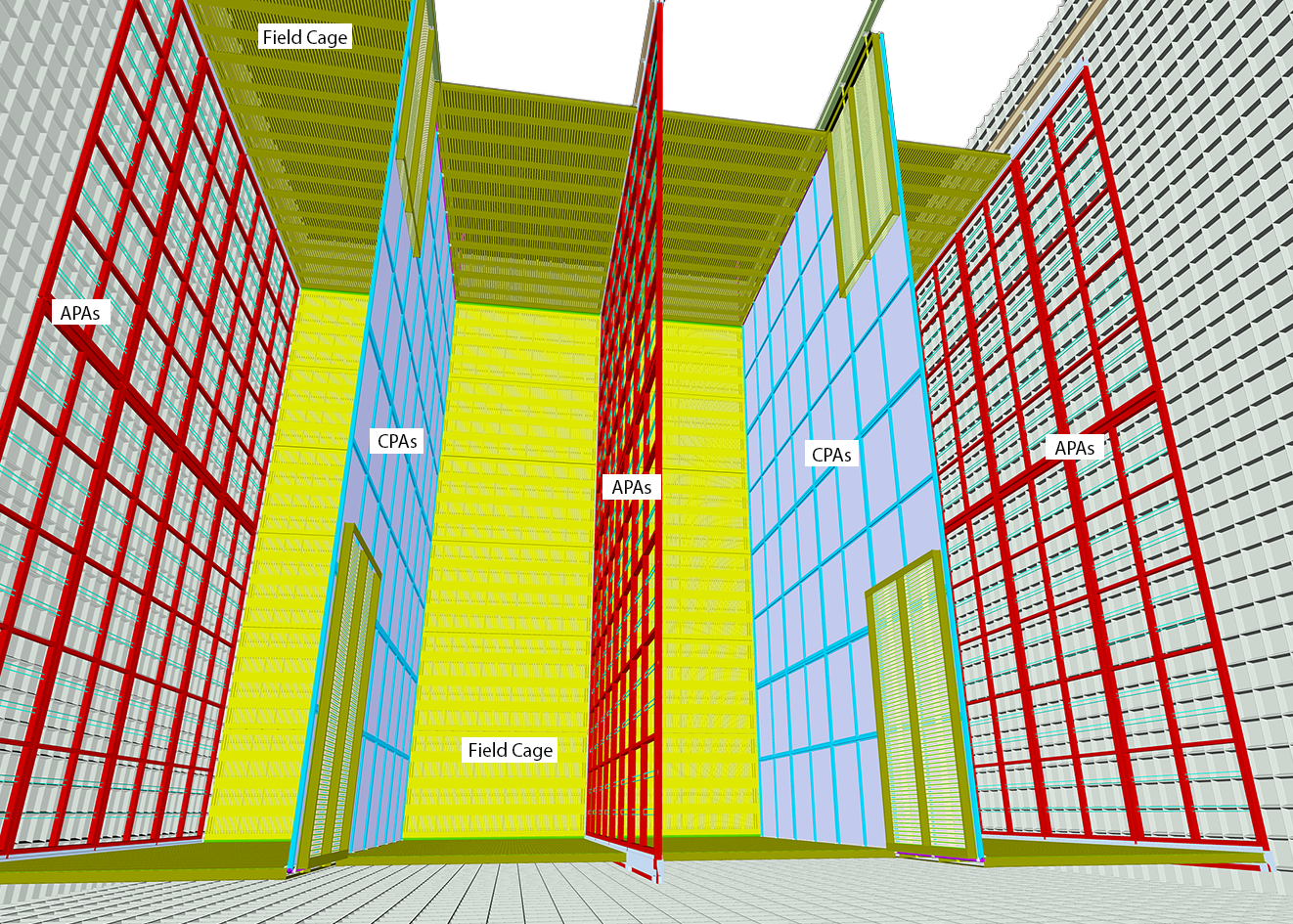}
  \caption{A view of the partially installed TPC inside the membrane cryostat. The APAs are shown in red, CPAs are in
    cyan, field-cage modules in yellow/green. Some of the field-cage modules are in their folded position against the
    cathode to provide aisle access during installation. Figure from \cite{cdrv4}.}
  \label{fig:tpc}
\end{figure}

The DUNE far detector (FD) is described in detail in Volume 4 of the DUNE CDR. The reference design consists of four 10-kt
fiducial mass, single-phase LArTPC modules with integrated photon detection systems. The active volume of one of these
far detector modules is 12~m high, 14.5~m wide, and 58~m long; this is instrumented with 150 anode-plane assemblies (APAs),
each of which has 2560 sense wires arranged in three wire planes and 200 cathode plane assemblies (CPAs). The TPC is
located inside a cryostat vessel which also contains field-cage modules to enclose the four open sides between the anode
and cathode planes. Figure \ref{fig:tpc} is a schematic showing the partially-installed detector.

\begin{figure}[!htpb]
  \centering
  \includegraphics[width=0.4\textwidth]{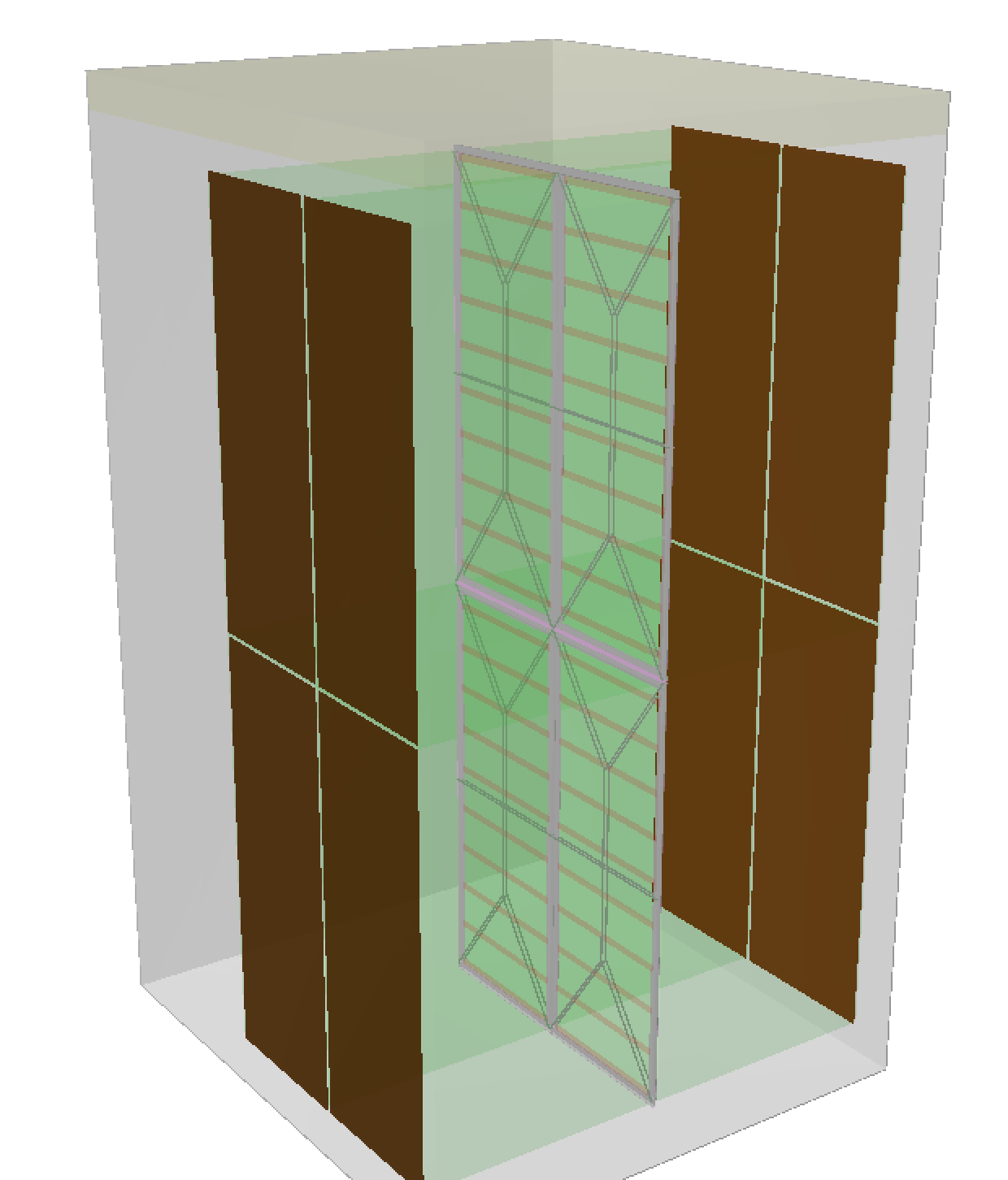}
    \includegraphics[width=0.4\textwidth]{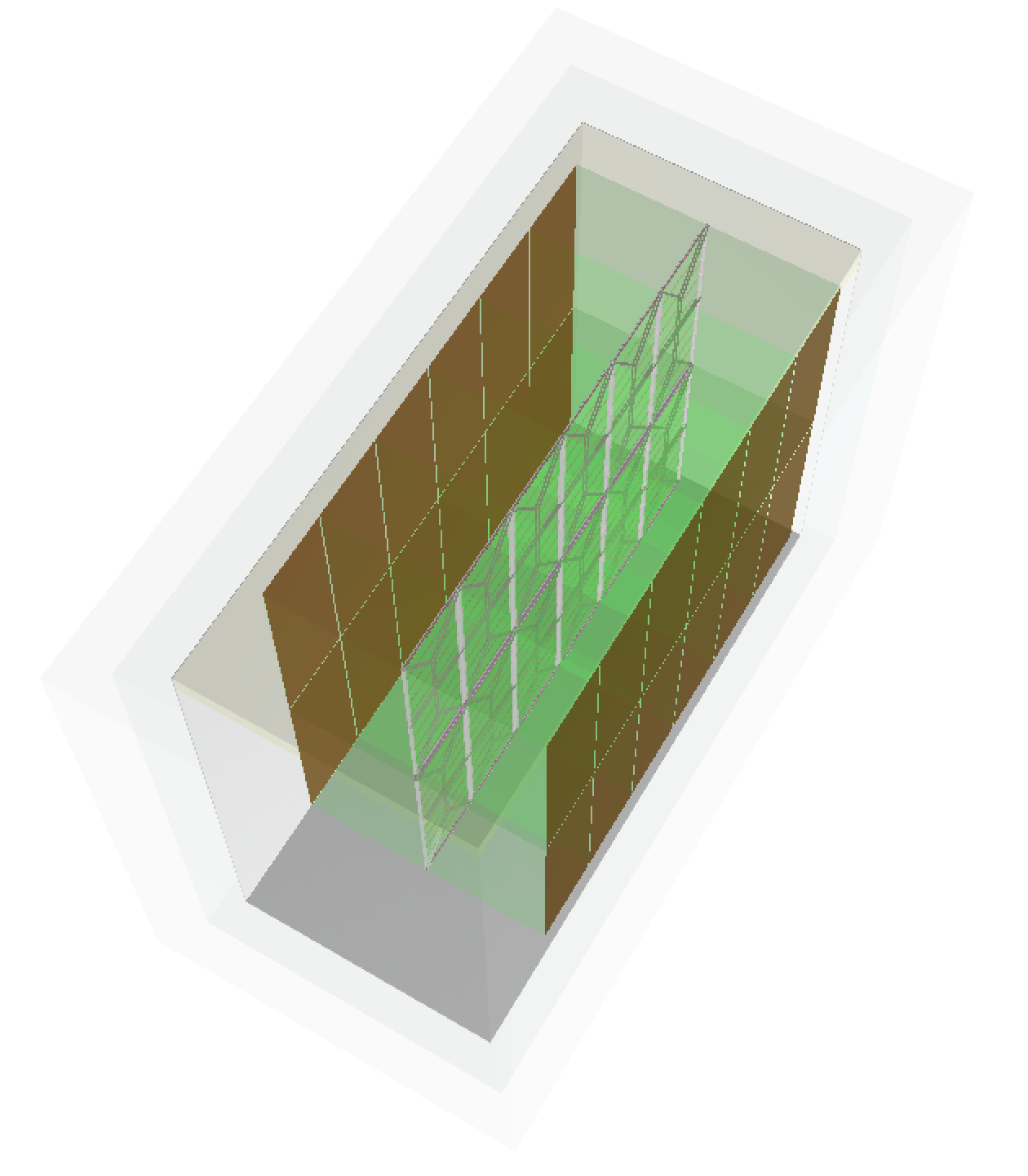}
    \caption{Visualizations of the two GDML configurations provided. The left image shows the smallest, 4-APA, workspace geometry.
      The right image shows the larger, 12-APA, workspace geometry, which provides more depth in the beam direction. The cathode
      planes are shown in brown and the APA frames are shown in gray.}
  \label{fig:geom_vis}
\end{figure}

Simulating the nearly 400k channels in a single FD cryostat is extremely costly, so most simulation studies are performed in a
smaller “workspace geometry” consisting of only a few APAs. The smallest (dune10kt\_v1\_workspace.gdml) is shown in the left image
of Fig.~\ref{fig:geom_vis} and consists of 4 APAs in
the center of the cryostat, two stacked vertically by two end-to-end, with 4 corresponding CPAs at either end of opposing drift volumes.
One feature of the APA is that one set of APA channels reads out the volume on both sides, with the opposite drift directions.
Providing an active volume 12 m tall, 7 m wide from CPA to APA to CPA, and 4.6 m long in the beam direction, this is the smallest geometry
which can still support studies involving gaps between vertically stacked and longitudinally adjacent APAs. Studies that rely on muon
versus hadron track length, however, need to use the next largest geometry (dune10kt\_v1\_1x2x6.gdml), shown in the right image of
Fig.~\ref{fig:geom_vis}, which has six APAs in the beam direction, providing 14 m of active LAr in the beam direction.
One feature of these workspace configurations is that the CPAs are on the outside so that the drift volume on both sides of the APA can be used,
whereas the actual FD design places the APAs on the outside, with the volume on the outer side not active. The GDML files are overwhelmingly
dominated by the wire description, so versions of the files that do not include the wires are also included. These ``nowires'' files are
especially useful for more efficient Geant4 tracking and for stand-along Geant4 studies, as LArSoft is needed to map sense wires to
the proper channels.

The GDML files for the two FD workspace geometries described here, with and without the APA sense wires, are provided in the ancillary files
in directory DUNE\_GDML/. These geometry descriptions may be used in conjunction with LArSoft\cite{larsoft} to perform a
Monte Carlo simulation of the DUNE far detector. Note that the full DUNE far detector simulation is under development and that this simulation
was not used to produce the sensitivities presented in the DUNE CDR.

\section{Summary}
The results of simulations of the LBNF neutrino beamline and a parameterized Fast Monte Carlo of the DUNE Far Detector
are provided to facilitate phenomenological studies of DUNE physics sensitivity. GDML files for simulation of the
DUNE single-phase far detector for use in LArSoft simulations are also provided. The DUNE collaboration welcomes
those interested in studying DUNE to join the collaboration or to use these configurations independently. Discussion of
of any results with the DUNE collaboration, either as a member or a guest, is encouraged. The collaboration requests
that any results making use of the ancillary files reference this arXiv posting and Volume 2 of the DUNE CDR\cite{cdrv2}.

\bibliographystyle{h-physrev}
\bibliography{cdr_configs_bib}

\end{document}